# Remote quantum states in curved spacetime

CHARLES FRANCIS[1]

*Jesus College, Cambridge*

It is seen that issues of unitarity raised by the evolution of the wave function in curved spacetime can be resolved by describing the evolution of quantum states in Minkowski tangent space. The treatment adheres closely to the orthodox interpretation that the wave function is not physical but is part of a mathematical method for the calculation of probabilities of measurement results. Minkowski tangent space refers to a non-physical configuration space used to describe quantum mechanics. The *teleconnection* is defined between Hilbert spaces at different points in spacetime motivated by arguments from the probability interpretation. The teleconnection is analogous to a connection between vector spaces and reduces to the Levi-Civita connection in the limit of near initial and final measurements. Predictions for quantum theory in curved spacetime agree with those of classical general relativity

## 1 Introduction

### 1.1 Context and terminology

It is often suggested that general relativity and quantum mechanics are not compatible, but in this paper I will show consistency between the classical theory of general relativity and the evolution of the wave function in Minkowski tangent space, and that evolution in Minkowski tangent space is required by the probability interpretation, where Minkowski tangent space is a non-physical configuration space determined from the probabilistic formulae underlying quantum theory. Any inconsistency does not lie in the *mathematical* structures or the predictions of quantum mechanics and general relativity, but in the *interpretations* that general relativity describes a substantive manifold and that quantum mechanics describes some form of ontic wave or field. The present development is in line with the orthodox (Dirac-von Neumann) interpretation of quantum mechanics, that it only makes sense to talk of measured values when a measurement is actually done, or when the outcome of a measurement can be predicted with certainty, that quantum mechanics describes knowledge of reality, not reality itself, and with von

---

1. C.E.H.Francis.75@cantab.net

Neumann's (1932) treatment in which the mathematical structure of quantum mechanics is established from the probabilistic results of measurements.

**Minkowski tangent space** will here mean an affine space tangent to the manifold and with global Minkowski metric, using coordinates defined such that the speed of light is unity. Tangent space is sometimes taken to mean the space of vectors defined at each point of a manifold. Since any vector space is an affine space over itself, the terminology is consistent. I will refer to Minkowski tangent space in order to emphasise the affine property, in keeping with the traditional meaning of a tangent (literally "touching") in Euclidean geometry and as used in early treatments of differential geometry where a manifold of dimension $n$ was typically embedded in a flat space, $\mathbb{R}^m$, of dimension $m > n$; tangent space was a flat subspace of $\mathbb{R}^m$ of the same dimension, $n$, as the manifold (e.g., Dirac 1975). Higher dimensional embedding is not used here. It is shown in appendix A that parallel displacement in Minkowski tangent space can be used to determine the Levi-Civita connection, following an argument based on that used by Dirac, but without using higher dimensions.

The **orthodox interpretation** is related to the Copenhagen interpretation, but is distinguished from it because it refers only to the probability interpretation and contains no notion of complementarity (Bub, 1997). It can be summarised in the words of Dirac (1958) "*In the general case we cannot speak of an observable having a value for a particular state, but we can ... speak of the probability of its having a specified value for the state, meaning the probability of this specified value being obtained when one makes a measurement of the observable*". We may infer that a precise value of position only exists when a measurement of position is performed or has certain outcome, so that we can only talk about where a particle is found in measurement, not where it is in space. In keeping with this interpretation, it will here be assumed that the spacetime manifold has meaning only in so far as it describes positions of matter which can actually be determined, and that these positions describe relationships between matter and reference matter, not relationships between matter and substantive spacetime (see appendix B).

In the orthodox interpretation, states in quantum mechanics can be identified with statements about possible results of measurements made by an observer with a clock at the origin, including statements in the future tense and statements in the subjunctive, concerning measurements which are not actually performed, though in principle might be performed (Francis 2013, 2015). As such, it contradicts naive realism by asserting that there is no direct correspondence between physical reality and statements which can be made about reality. Hilbert space concerns an observer's knowledge of reality, not reality itself. The wave function is a mathematical device used in the calculation of probabilities, not an actual property of reality. It should then be understood that Hilbert space applies to the measurements by an observer at particular position, not to a wave function on a space-like hypersurface in a substantive spacetime.

A **Schrödinger equation** will mean an equation with the form $\partial_0 f(t, x) = -iHf(t, x)$ (C.13), which can be relativistic (appendix C). Indeed, Dirac referred to the Dirac equation as a Schrödinger equation, and found the Dirac equation by requiring an equation of this form. For clarity I suppress spin, which has no impact on the treatment here. For simplicity, I treat only the application to photons. To apply the definition of the teleconnection to other particles it is necessary to use a relativistic wave function obey-ing the mass shell condition (or the Klein-Gordon equation). A non-relativistic wave function will give nonsensical results because of the arbitrary zero of potential energy.

This treatment uses relativistic wave functions rather than quantum fields. Haag's theorem (1955) effectively prohibits the definition of "interacting fields" (also Hall & Wightman, 1957). Nonetheless, quantum fields are required for theories of interactions between fundamental particles, and are related to wave functions using Fock space (Francis 2013), as is also done in causal perturbation theory (e.g., Scharf 1989) and lattice field theory (e.g., Montvay and Münster 1994). In these accounts quantum fields are operators (or operator valued distributions) on the Fock space of non-interacting particles. Interactions are modelled by compositions of quantum field operators. In general relativity, geodesic motions are inertial; interactions are not involved. Wave functions (in a multiparticle Fock space) are appropriate for the analysis of the inertial motions of non-interacting particles and are consequently more suited to a general relativistic treatment in which inertial motions are indicated by the equivalence principle. The results of this paper are then automatically be carried through to quantum field theory, when interactions are introduced.

Eduard Prugovecki (1994, 1995, 1996) has also considered quantum propagation in curved spacetime represented by a globally hyperbolic Lorentzian manifold $(M, g)$, using parallel transport governed by a quantum connection obtained by extending the Levi-Civita connection from the Lorentz frame bundle to the Poincaré frame bundle over $(M, g)$, finding that this quantum connection gives rise to infinitesimal parallel transport coinciding with special relativistic quantum evolution. However, his path-integral formulation carries a much greater overhead of mathematical complexity than the treatment of this paper. Consequently the consistency of the quantum treatment with classical gravity is less transparent.

*1.2 Paper structure*

Section 2 will describe difficulties concerning the formulation of quantum mechanics in curved space time and clarify the requirement for quantum evolution of the wave function in Minkowski tangent space. Although classical wave motions can be modeled in curved spacetime and treatments of aspects of quantum field theory in curved spacetime have been given (e.g. Fulling 1989, Wald 1994, Parker & Toms, 2009), I will specifically consider requirements of unitarity in quantum mechanics arising from Stone's theorem. On an intuitive level, the requirement is that the probability represented by the wave function sums to unity on a synchronous slice. But the path dependency of proper time means that this does not remain true for wave evolution in a general curved spacetime. The formal proof from Stone's theorem makes it clear that evolution of the wave function is determined using non-physical Minkowski metric, not the physical metric of spacetime.

Section 3.1 will describe parallel displacement between initial and final states. In section 3.2 it will be seen that this leads to a consistent formulation of quantum theory in a gravitational field. The formal definition of the teleconnection (3.1.5) projects the evolved wave function back to the spacetime manifold and makes explicit the full consistency between quantum theory and general relativity. This is possible because in the orthodox interpretation of quantum theory position exists only in measurement. In contrast to a classical field, wave functions in quantum mechanics are not observable, even in principle. In the orthodox interpretation, the wave function is merely an expression of probability theory and has no other ontological meaning (appendix B). The present development does not assume a substantivalist manifold and is consistent with particle QED as described informally by Feynman (1985) and formally in Francis (2013) in which light consists of photons and the classical electromagnetic field is the expectation of a Hermitian operator.

Conclusions are described in section 3.4. Appendices have been included to show that parallel displacement in Minkowski tangent space gives the Levi-Civita connection (appendix A), to clarify what is meant by the orthodox interpretation of quantum mechanics (appendix B), to show that the probability interpretation requires the solution of the Schrödinger equation in Minkowski tangent space (appendix C), that the Schrödinger equation for a non-interacting particle leads to geodesic motion in the classical correspondence (appendix D).

## 2 A remote connection for quantum states

### 2.1 The requirement for a remote connection

Quantum mechanics enables us to calculate probabilities for the results of a final measurement, given information taken from measurement and described in the initial state. It is generally assumed that the initial and final measurements can be described in a single reference frame and treated using Minkowski metric. By **reference frame** I do not mean coordinate system, but rather the chosen matter from which a coordinate system may be determined in practice by physical measurement, as in, e.g., "the Earth frame", "the lab frame", or "the frame of the fixed stars". In special relativity, time is defined from a clock local to an observer. It is assumed that distant clocks can be calibrated using Einstein's (1905) synchronisation procedure. This method can be used in the presence of a gravitational field, for example to calibrate clocks on GPS satellites to a master clock on earth. In this case we can treat both clocks within a single reference frame. However, at sufficient distances, there is no physical mechanism by which clocks can be calibrated. There is then a problem of principle with the assumption that physics can be described using a single time parameter, or (equivalently) in a single reference frame. The teleconnection (section 3.3) defines a mapping between Hilbert spaces, preserving unitarity and respecting the principle of general relativity, according to which Hilbert space is defined relative to local matter.

It is usually the case that the physical scale of quantum effects is small, but there is no theoretical limit on the distance of the propagation of a wave function. In general, even when synchronisation is possible, clocks at different positions do not remain synchronised because of the effects of gravity. Consequently questions are raised concerning the formulation of quantum mechanics in curved spacetime. The equivalence principle guarantees that every point of a Lorentzian manifold has a Minkowski tangent space, meaning that Minkowski metric can be used locally and that the Christoffel symbols vanish at that point. Physically this means that an observer can define inertial coordinates within a neighbourhood, for example using light triangulation (the radar method), such that the physical metric, *g*, is equal to Minkowski metric, *h*, at the origin. Then a Minkowski tangent space can be defined using the same coordinates together with constant, non-physical, Minkowski metric, *h*. A connection defines infinitesimal parallel displacement, such that vectors are translated small distances in Minkowski tangent space and projected back onto the manifold. In the real world, this means that objects can be translated through small distances within a neighbourhood such that differences between physical measurement and corresponding calculations in Minkowski tangent space are negligible.

In general relativity, spacetime vectors are defined at each point in tangent space. Similarly, kets are vectors in Hilbert space at a given location, determined by the measurement apparatus of an observer. In the orthodox interpretation of quantum mechanics it is meaningless to talk of measured quantities, like position, between measurements. Physically meaningful quantities are found only when a measurement is actually performed or when the result of a measurement is deterministic. Strictly the properties of a manifold are only present in measured quantities (including deterministic results of measurements which are not actually performed). The teleconnection for Hilbert space is not defined by infinitesimal parallel displacement of kets to points where no measurement takes place, but between Hilbert spaces representing the initial and final measurements for an intervening quantum process, regardless of the origins at which those Hilbert spaces are defined.

Between measurements, the probability interpretation imposes wave evolution as described by the solution of the Schrödinger equation with flat space metric (section 2.2 and appendix C). Thus, there is a deep and fundamental incompatibility between the probability interpretation of quantum mechanics and the solution of a wave equation in curved spacetime. The probability interpretation has a fundamental role in quantum mechanics, and it does not appear that it can, or should, be altered to allow wave evolution in curved space time. This paper proposes a resolution by postulating that the Schrödinger equation is solved in Minkowski tangent space and the resulting state is projected back into spacetime at the time of the collapse of the wave function. It will follow that classical curved spacetime appears as the envelope of Minkowski tangent spaces used to describe quantum physics (figure 1).

Thus, Hilbert space, $\mathbb{H}(x)$, will be defined at an origin *x* by an observer. The solution of the Schrödinger equation, (2.2.2), describes the evolution of components of kets in Minkowski tangent space, not the evolution of a wave on physical spacetime. At a second measurement, at *y*, the state is described by a ket in another Hilbert space, $\mathbb{H}(y)$, defined with an origin at *y*. The *teleconnection*, described in section 3, is defined as a sesquilinear product between $\mathbb{H}(x)$ and $\mathbb{H}(y)$, $\mathbb{H}(y) \times \mathbb{H}(x) \to \mathbb{C}$, and determines how the collapse of the wave function projects the evolved state in $\mathbb{H}(x)$ to a state in $\mathbb{H}(y)$. In the limit of an infinitesimal displacement from *x* to *y* the result of the teleconnection is identical to parallel displacement of momentum under the Levi-Civita connection.

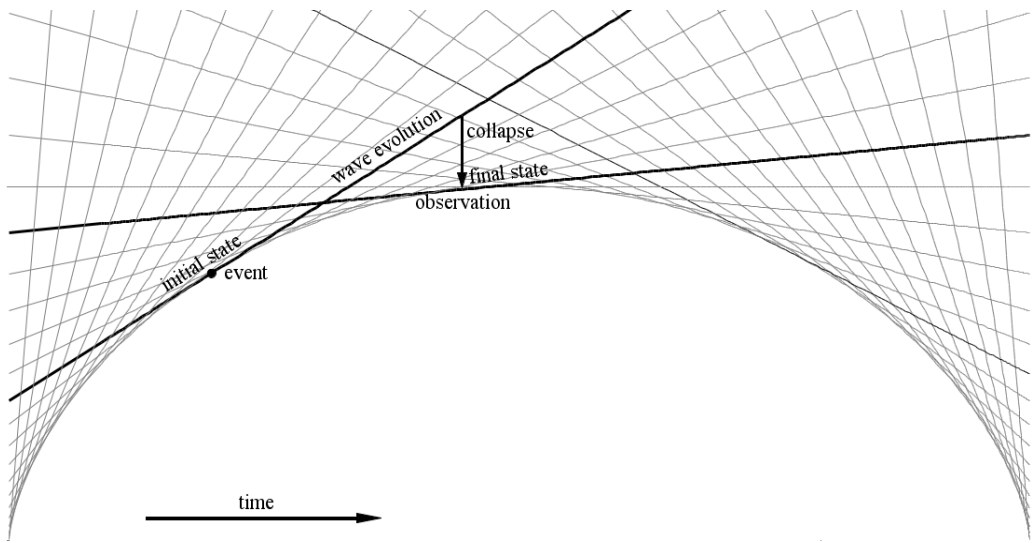

**Figure 1:** A smooth manifold may be defined as the envelope of a family of flat (or affine) tangent spaces. The evolution of the wave function from an initial state is described within Minkowski tangent space. The teleconnection then projects the evolved state to the final state, described in a different Minkowski tangent space, at the time of collapse of the wave function. The time coordinate on the horizontal axis can be regarded as cosmic time, as distinct from observer time coordinate which would be represented within the manifold. The cycloidal form shown here applies to a closed Friedmann model with no cosmological constant, but an envelope can be described for any model.

### 2.2 The inconsistency of curvature with wave mechanics

According to the orthodox, or Dirac-von Neumann, interpretation of quantum mechanics a probability amplitude is not conceived as describing a fundamental property of matter, but rather describes an observer's knowledge of measurement results; it does not describe a physical wave, but is a means of calculating the probability for the outcome of an experiment. Since measurement results are just numerical values, and measurements of quantum states do not admit measurements of distances remote from the observer, the physical spacetime metric has no role in Hilbert space. It is postulated that the evolution of states is given by the time evolution operator, $U(t, t_0): \mathbb{H} \to \mathbb{H}$, such that if at time $t_0$ the ket is $|f(t_0)\rangle$, then the ket at time $t$ is $|f(t)\rangle = U(t, t_0)|f(t_0)\rangle$. As described in appendix C, the probability interpretation requires that $U$ is unitary. Then it follows from Stone's (1932) theorem that there exists a Hermitian operator $H$, the Hamiltonian, such that $\dot{U}(t) = -iHU(t)$ which has solution, $U(t) = e^{-iHt}$. The Schrödinger equation follows immediately.

For an initial state $|f\rangle$ at time $x^0 = t_0$ described by a wave function $f(x) = \langle x | f \rangle$, the momentum space wave function $\langle p | f \rangle$ is given at time $x^0 = t_0$ by

$$\langle p | f \rangle = \left(\tfrac{1}{2\pi}\right)^{3/2} \int d^3 x \langle x | f \rangle \, e^{-i \mathbf{x} \cdot \mathbf{p}}, \qquad (2.2.1)$$

where the dot product uses the Euclidean metric at constant time. The Euclidean metric in (2.2.1) is not the physical metric and merely defines the momentum space wave func-tion as a transform of the initial condition. For a non-interacting particle, $H = E$ = const. The general solution for times $x^0 = t \geq t_0$ is

$$f(x) = \langle x|f \rangle = \left(\frac{1}{2\pi}\right)^{3/2} \int d^3p \langle p|f \rangle \, e^{-ix \cdot p} \tag{2.2.2}$$

where $\langle p|f \rangle$ is constant in time. The dot product, $x \cdot p$, appearing in the exponent under the integrand in (2.2.2) uses Minkowski metric. It is important for the development in this paper to recognise that this has arisen from the solution of a differential equation, *not* from the physical metric of spacetime. Minkowski metric is here a property of the Hilbert space $\mathbb{H}(A)$ defined by an observer, Alf, at an origin A; it is strictly a tensor value, not a tensor field. Equivalently, the solution (2.2.2) is found in Minkowski tangent space such that the coordinate speed of light is unity, and plane wave properties are preserved. (2.2.2) expresses Newton's first law for a non-interacting particle in Minkowski tangent space. Geodesic motion follows for non-interacting classical particles (appendix D).

## 3 The Teleconnection

### 3.1 Parallel displacement in Minkowski tangent space

An inertial observer, Alf at A, defines normal coordinates such that the radial speed of light is unity. Restricting to the radial direction, in normal coordinates defined with unit lightspeed, the line element has the form

$$ds^2 = k^2 d\tau^2 - k^2 d\rho^2, \tag{3.1.1}$$

where $k(A) = 1$. At each point in normal coordinates it is required to specify only a scale factor to determine a Minkowski tangent space. It is always possible to define normal coordinates within a neighbourhood of an inertial observer (for example using the radar method). Friedmann cosmologies can also be described using normal coordinates,

$$ds^2 = a^2(\tau)[d\tau^2 - d\rho^2 - f^2(\rho)(d\theta^2 + \sin^2\theta d\phi^2)] \tag{3.1.2}$$

Alf defines Hilbert space, $\mathbb{H}(A)$, at origin A, using locally Minkowski coordinates. The solution of the Schrödinger equation for a non-interacting particle (2.2.2) describes the parallel displacement of momentum $p$ in Minkowski tangent space, meaning that components of momentum in normal coordinates are constant.

$$p(B) = p(A). \tag{3.1.3}$$

A second inertial observer, Beth, at B, similarly defines Minkowski tangent space at B. Without loss of generality, at the time of Beth's measurement, Beth is stationary in Alf's coordinates (since Lorentz transformation can be applied to coordinate system of a moving observer). Normal coordinates, (3.1.1), are transformed to locally Minkowski coordinates at B using

$$(t, r) = k_B(\tau, \rho), \tag{3.1.4}$$

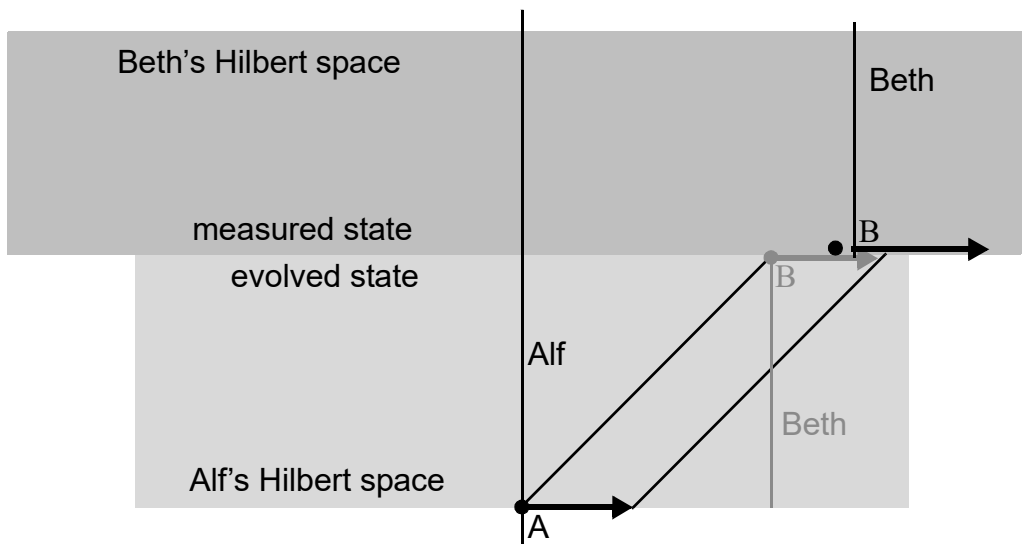

**Figure 2:** Parallel displacement of a wavelength of light in Minkowski tangent space. In Alf's coordinates, proper period at A is $\lambda$. At B it will be $k\lambda$ after enlarging Alf's tangent space to match Beth's. In a gravitational field, $k$ describes gravitational redshift. In a Friedmann cosmology, $k = a_B / a_A$ describes cosmological redshift.

where $k_B = k(B)$.

In a measurement by Beth at B, the state is described by a ket in another Hilbert space, $\mathbb{H}(B)$, defined by Beth with an origin at B.

According to the general principle of relativity, the distance scales of Alf's and Beth's coordinate systems are identical (figure 2); both are defined locally using the standard definition of the second and with speed of light equal to unity. Alf and Beth have defined Hilbert spaces $\mathbb{H}(A)$ and $\mathbb{H}(B)$ using bases of position states defined with respect to their own local distance scales. Unitarity requires an isomorphism between $\mathbb{H}(A)$ and $\mathbb{H}(B)$, necessitating an enlargement such that the product at time $t_B$ between the *evolved* states in $\mathbb{H}(A)$ and the *observed* states in $\mathbb{H}(B)$ restores position. The value of normal coordinates is that they allow us to replace disjoint tangent spaces, seen in figure 2, with a continuous space and to see the consistency of the teleconnection with curved spacetime.

Thus, for Hilbert spaces, $\mathbb{H}(A)$ and $\mathbb{H}(B)$, defined at A and B respectively, the **teleconnection** is defined by the product, $H(B) \times H(A) \to \mathbb{C}$, such that for momentum $p$ at A, and momentum $q$ at B,

$$\langle q | p \rangle = \left(\frac{1}{2\pi}\right)^3 \int d^3x \, e^{-ix \cdot (k_B p - q)} = \delta(k_B p - q), \tag{3.1.5}$$

where the axes of Minkowski tangent space at A and B are aligned and $p$ is translated to B using (3.1.3). Since (3.1.5) describes parallel displacement of momentum in tangent space, it follows immediately that for small displacements it is equivalent to the Levi-Civita connection (appendix A). Geodesic motion follows in the classical correspondence, for systems in which measurement outcomes can be continuously predicted with certainty (appendix D).

## 3.2 Gravitational redshift

An inertial observer, Alf, defines normal coordinates for a static geometry using time $t$, determined from a clock at origin, A. Ignoring expansion, and restricting to the radial direction, in normal coordinates defined with unit lightspeed, the line element has the form (3.1.1). $\mathbb{H}(A)$ is simply the Hilbert space of standard quantum mechanics, and the usual relationship between energy-momentum and frequency-wavelength holds locally to Alf. Let coordinate time at A be $\tau_A$ and at B be $\tau_B$. $k(A) = 1$. Let $k_B = k(B)$. A photon of energy $E$ at A has period $\lambda = 1/E$ in locally Minkowski coordinates at A. On translation to B in Minkowski tangent space its period is still $\lambda$ in normal coordinates, or $k_B \lambda$ in locally Minkowski coordinates. Then the energy $E_B$ of the photon at B is given by

$$E_B = \frac{1}{k_B} E_A, \qquad (3.2.1)$$

in agreement with standard general relativity. Indeed, because light-like geodesics are drawn at 45° in normal coordinates, the argument is identical in these coordinates to the classical argument using geodesic motion for the initial and final point defining one wavelength of light. Clearly the choice of a particular coordinate system makes no difference to the result. The sole difference is that a quantum wave function has been used rather than a classical wave.

## 3.3 Expansion

Consider a Friedmann cosmology with metric

$$ds^2 = a^2(d\tau^2 - d\rho^2 - f^2 d\Omega^2) \qquad (3.3.1)$$

where $f = \sin\rho, \rho, \sinh\rho$ for space with positive, zero or negative curvature. A photon travels from Alf at A to Beth at B. Let the expansion factor be $a_A$ at emission and $a_B$ at observation. Using the substitutions $t_A = a_A \tau$, $r_A = a_A \rho$, $t_B = a_B \tau$, and $r_B = a_B \rho$ we obtain locally Minkowski coordinates for Alf and Beth at the time of emission and the time of detection of the photon

$$ds^2 = \frac{a^2}{a_A^2}(dt_A^2 - dr_A^2 - a_A^2 f^2 d\Omega^2) \qquad (3.3.2)$$

and

$$ds^2 = \frac{a^2}{a_B^2}(dt_B^2 - dr_B^2 - a_B^2 f^2 d\Omega^2) \qquad (3.3.3)$$

respectively. In figure 2 Alf's position, A, has been chosen as the origin for both coordinate systems. It is assumed that Alf and Beth have constant position in comoving coordinates. The universe expands by factor $a_B/a_A$ between time of emission and time of detection. As defined in $\mathbb{H}(A)$, a wave function evolves as a plane wave in Minkowski tangent space at A. Following evolution in tangent space, unitarity requires an isomorphism between $\mathbb{H}(A)$ and $\mathbb{H}(B)$, necessitating an enlargement by a factor $a_B/a_A$, being equal to cosmological redshift. More generally we can treat geometries with both cosmological and gravitational redshift in normal coordinates.

### 3.4 Conclusion

It has been shown that there is complete consistency between general relativity and quantum mechanics when it is accepted that wave evolution takes place in Minkowski tangent space, not in curved spacetime. It has been seen that evolution in Minkowski tangent space is required by the probability interpretation and follows from the application of Stone's theorem. The continuous manifold of standard general relativity is mathematically described as the envelope of Minkowski tangent spaces used in the quantum theory, where Minkowski tangent space is a non-physical configuration space required for a rigorous treatment of the probabilistic structure of quantum mechanics. The construction is conceptually, but not technically, difficult, because it requires that we cease to think of space or spacetime as an arena into which matter may be placed. This difficulty may be compared to the conceptual problems many people have in relativity, where Newtonian space and time are replaced with spacetime. In its logical conclusion, relativity of measurement incorporates both relativity and quantum mechanics in one system, in which spacetime is no longer a fundamental prior of reality. It is neither conventional nor easy to think in such terms, but we must be guided by mathematical consistency and empirical fact, rather than make judgements influenced by conflict with preconception.

The teleconnection has been defined as a sesquilinear product between Hilbert spaces defined at different points in space time, and plays a role analogous to a connection between vector spaces. The teleconnection reduces to the Levi-Civita connection for small displacements. It is clear that gravitational redshifts for quantum states under the teleconnection are identical to gravitational redshifts for classical electromagnetic radiation, because the quantum treatment is identical to a classical treatment using normal coordinates, and because results are not coordinate dependent.

# Appendices

## Appendix A   Calculation of the Levi-Civita Connection

In modern treatments the Christoffel symbols typically define the connection and are then used to define parallel transport. This is a reversal of the traditional approach in which the connection was defined from infinitesimal parallel displacement in Minkowski tangent space. The change in approach has been motivated because of the difficulty in giving a satisfactory justification of a higher dimensional space in which Minkowski tangent space was embedded. In this treatment Minkowski tangent space is defined as in section 2.1. Embedding in higher dimensions has not been used. Parallel displacement in Minkowski tangent space applies in the quantum domain, and reduces to parallel transport in the limit of small parallel displacement. This appendix is included for completeness and shows the Levi-Civita connection from infinitesimal parallel displacement. It closely follows the derivation given by Dirac (1975), excepting that it does not depend on embedding in higher dimensions. I believe this has some advantage of conceptual simplicity over more usual approaches. Clearly the argument can be generalised to other manifolds using an affine tangent space. This paper is only concerned with the spacetime manifold.

### *A1   Christoffel Symbols*

Consider a region in which the metric field is $g_{ab}$.

**Definition:** Christoffel symbols of the first kind:

$$\Gamma_{abc} = (g_{ab,c} + g_{ac,b} - g_{cb,a})/2 \,.$$

**Definition:** Christoffel symbols of the second kind:

$$\Gamma^a_{\ bc} = g^{ad}\Gamma_{dbc}\,.$$

Clearly Christoffel symbols are symmetric in the last two suffixes, and satisfy

$$\Gamma_{abc} + \Gamma_{bac} = g_{ab,c}.$$

The purpose of Christoffel symbols is to enable us to calculate the effect of parallel displacement without reference to a non-physical tangent space. Tangent space at $x$ is defined with non-physical Minkowski metric $h$ using primed coordinates. We have

$$g_{ab}(x) = k_a^{m'}(x) k_b^{n'}(x) h_{m'n'}. \tag{A1.1}$$

$$g_{ab}(x+dx) = k_a^{m'}(x+dx) k_b^{n'}(x+dx) h_{m'n'} + O(\max(dx^i)). \tag{A1.2}$$

$$g_{ab,c}(x) = k_{a,c}^{m'}(x) k_b^{n'}(x) h_{m'n'} + k_a^{m'}(x) k_{b,c}^{n'}(x) h_{m'n'}. \tag{A1.3}$$

Interchange $a$ & $c$ and $b$ & $c$ in (A1.3).

$$g_{cb,a} = k_{c,a}^{m'} k_b^{n'} h_{m'n'} + k_c^{m'} k_{b,a}^{n'} h_{m'n'},$$

$$g_{ac,b} = k_{a,b}^{m'} k_c^{n'} h_{m'n'} + k_a^{m'} k_{c,b}^{n'} h_{m'n'}.$$

Then, noting that, from Clairaut's Theorem, the partial derivative of the transformation matrix is symmetrical in its lower indices, $k_{c,b}^{m'} = k_{b,c}^{m'}$.

$$\Gamma_{abc} = (g_{ab,c} + g_{ac,b} - g_{cb,a})/2 = k_a^{m'} k_{b,c}^{n'} h_{m'n'}. \tag{A1.4}$$

It should be noted that this obtains only at the point, $x$, where Minkowski tangent space touches the spacetime manifold. (A1.1) means only that the manifold is locally Minkowski. Since a manifold and a Minkowski tangent space are different spaces, it cannot be reinterpreted as a global coordinate transformation (which can only apply to the same space and would mean that the manifold is flat). This may be compare the statement that the gradient of a curve at a point is equal to the gradient of the tangent at that point. Clearly the gradient of a curve is not everywhere equal to the gradient of the tangent, since otherwise the curve would be a straight line.

## A2   Parallel Displacement

Parallel displacement of a vector $p$ from $x$ to $x+dx$ in Minkowski tangent space keeps the primed components constant, $p^{m'}(x+dx) = p^{m'}(x)$. Multiply by $k_b^{n'}(x+dx) h_{m'n'}$, to lower the index and convert to unprimed coordinates.

$$p_b(x+dx) = p^{m'}(x) k_b^{n'}(x+dx) h_{m'n'} + O(\max(dx^i)).$$

$$p_b(x+dx) = p_b(x) + p^{m'}(x) k_{b,c}^{n'}(x) dx^c h_{m'n'} + O(\max(dx^i)).$$

Then, ignoring terms $O(\max(dx^i))$,

$$dp_b = p^{m'}(x) k_{b,c}^{n'}(x) dx^c h_{m'n'} = p^a(x) k_a^{m'}(x) k_{b,c}^{n'}(x) dx^c h_{m'n'}.$$

Substituting the Christoffel symbol eliminates the dependency on tangent space.

$$dp_b = \Gamma_{abc} p^a dx^c.$$

Raise and lower indices to find the standard formula for infinitesimal parallel displacement referring to covariant components.

$$dp_b = \Gamma^a_{bc} p_a dx^c.$$

For a second vector $q$, $p \cdot q$ is invariant.

$$d(p_b q^b) = 0.$$

$$dp_b q^b + p_b dq^b = 0.$$

$$p_a \Gamma^a_{bc} dx^c q^b + p_a dq^a = 0.$$

Since this is true for all $p_a$, we find the standard formula for infinitesimal parallel displacement referring to contravariant components,

$$dq^a = -\Gamma^a_{bc} q^b dx^c$$

### Appendix B  The orthodox interpretation

In his introduction to The Mathematical Theory of Relativity, Sir Arthur Stanley Eddington (1923) said *"A physical quantity is defined by the series of operations and calculations of which it is the result"*. Although written in the context of relativity, and prior to the complete formulation of quantum mechanics, Eddington's words, taken to their logical conclusion, strongly suggest the orthodox interpretation, in which, in the general case, physical quantities, such as position, cannot be said to exist when a measurement is not performed. The natural conclusion is that if position does not exist, then prior space does not exist.

A full treatment showing the necessity for Hilbert space from the statistical results of measurement was given by von Neumann (1932). For example, in the case of the Young's slits experiment, quantum mechanics does not make a statement about which slit the particle comes through, but it does make a statement about the probability for finding the particle at one or other slit if an experiment were to be done. Hilbert space can be treated as a language about an observer's knowledge of reality, not reality itself (Birkhoff and von Neumann, 1936). The wave function is then a mathematical device used in the calculation of probabilities, not a property or description of reality. In the case of Young's slits, the interpretation is that the property of position only exists as a consequence of interaction between the particle and its environment. We can calculate the probability that a particle would be found coming through either slit if an observation were carried out, but, in the absence of interaction between the particle and its environment, it is not meaningful to describe a spacial relationship between the particle and its environment, that the particle comes through one or other slit.

Standard quantum mechanics describes the probabilities of measurement results in a final state given knowledge of an initial state (derived from measurements and/or the processes by which the system was prepared). We assume, for convenience, that the final measurement is a measurement of position. We may define a complex amplitude with arbitrary phase by taking the square root of the probability distribution. We may then treat these amplitudes as members of a complex Hilbert space, and denote them by kets. Because position states are orthogonal, it follows immediately from the fact that measurements of different positions are mutually exclusive that the inner product on Hilbert space restores the probability distribution, and that, after normalisation, the superposition of amplitudes corresponds directly to the probability of obtaining one OR other result. Then it is seen that rather than start with amplitudes, as in many treatments, we may start with probabilities and define amplitudes using straightforward mathematical definitions, provided only that we can justify the Schrödinger equation.

The rationale for the Schrödinger equation is well established (e.g. Jauch 1968, Simon 1976), but even now it is little considered in standard textbooks of quantum mechanics and it is still probably known only to a relatively small number of experts in the mathematical foundations of quantum theory. I outline it here, and give proofs in appendix C. Between measurements there is no change in information. Then the result of the calculation of probability is not affected by the time at which it is calculated. Since phase is arbitrary, we may choose it to be continuous. So time evolution is modelled by a continuous operator valued function of time, $U(t, t_0)$, on Hilbert space, such that $|f(t)\rangle = U(t, t_0)|f(t_0)\rangle$. The probability interpretation requires that $U$ is linear and can be chosen to conserve the norm. It follows that $U$ is unitary and satisfies the conditions of Stone's theorem. Thus, there exists a Hermitian operator $H$, the Hamiltonian, such that $\dot{U}(t) = -iHU(t)$. This has solution $U(t) = e^{-iHt}$. The Schrödinger equation follows immediately. It is therefore seen that the Schrödinger equation is required by the general considerations of probability theory.

In the account described above, Hilbert space is constructed from standard probability theory, and contains no more information than probability theory. For example, since the states of an alive and dead cat are orthogonal in Hilbert space, their superposition is nothing more than a probabilistic statement that the cat may be alive or may be dead. Collapse has nothing to do with the behaviour of an ontic wave, but is merely a change in probability when information becomes available. The derivation of the Schrödinger equation by way of Stone's theorem relies on the condition that there is no change in available information between the initial and final measurement. Thus the fundamental distinction between quantum evolution and classical evolution is that the quantum theory describes an initial state and a final state with no intervening observation whereas continuous measurement is possible for classical evolution, at least in principle (continuous measurement may include calculation from known initial conditions using a determinist equation of motion).

This suggests that the difference between quantum and classical physics is that we must abandon the classical notion of substantive spacetime as $\mathbb{R}^4$. Although Pythagoras decreed that the universe is composed of pure number, and this concept found its way into physics by way of Newtonian space and time, in modern mathematics the concept of number is developed without reference to reality. The natural interpretation of quan-tum mechanics is that spacetime is emergent and may only be said to exist when it is created from the configuration of matter.

## Appendix C  Justification of the Schrödinger equation

The inner product allows us to calculate probabilities for the outcome of a measurement provided that we know the ket describing hypothetical measurement at the time of measurement. We assume that Hilbert space is spanned by position states, $|x\rangle$, and denote the position state $|x\rangle$ at time $t$ by $|t, x\rangle$. The probability interpretation requires continuous linear time evolution:

**Postulate:** If at time $t_0$ the ket is $|f(t_0)\rangle$, then the ket at time $t$ is given by a continuous linear **time evolution operator**, $U(t, t_0)\colon \mathbb{H} \to \mathbb{H}$, such that $|f(t)\rangle = U(t, t_0)|f(t_0)\rangle$.

Since local laws of physics are always the same, and $U$ does not depend on the state on which it acts, the form of the evolution operator for a time span $t$, $U(t) = U(t+t_0, t_0)$, does not depend on $t_0$. We require that the evolution in a span $t_1 + t_2$ is the same as the evolution in $t_1$ followed by the evolution in $t_2$, and is also equal to the evolution in $t_2$ followed by the evolution in $t_1$, $U(t_2)U(t_1) = U(t_2+t_1) = U(t_1)U(t_2)$. In a zero time span, there is no evolution. So, $U(0)$ does not change the state; $U(0) = 1$. Using negative $t$ reverses time evolution (put $t = t_1 = -t_2$); $U(-t) = U(t)^{-1}$.

**Theorem:** $U$ is unitary.

*Proof:* In the absence of further information, the result of the calculation of probability of a measurement result $g$ at time $t_2$ given an initial condition $f$ at time $t_1$ is not affected by the time at which it is calculated. Since states can be chosen to be normalised we may require that $U$ conserves the norm, i.e., for all $|g\rangle$,

$$\langle g|U^\dagger U|g\rangle = \langle g|g\rangle. \tag{C.1}$$

Applying this to $|g\rangle + |f\rangle$,

$$(\langle g| + \langle f|)U^\dagger U(|g\rangle + |f\rangle) = (\langle g| + \langle f|)(|g\rangle + |f\rangle). \tag{C.2}$$

By linearity of $U$,

$$(\langle g|U^\dagger + \langle f|U^\dagger)(U|g\rangle + U|f\rangle) = (\langle g| + \langle f|)(|g\rangle + |f\rangle). \tag{C.3}$$

By linearity of the inner product

$$\langle g|U^\dagger U|g\rangle + \langle g|U^\dagger U|f\rangle + \langle f|U^\dagger U|g\rangle + \langle f|U^\dagger U|f\rangle$$
$$= \langle g|g\rangle + \langle g|f\rangle + \langle f|g\rangle + \langle f|f\rangle$$

Thus, from (C.1),

$$\langle g|U^\dagger U|f\rangle + \langle f|U^\dagger U|g\rangle = \langle g|f\rangle + \langle f|g\rangle. \tag{C.4}$$

Similarly conservation of the norm of $|g\rangle + i|f\rangle$ gives

$$\langle g|U^\dagger U|f\rangle - \langle f|U^\dagger U|g\rangle = \langle g|f\rangle - \langle f|g\rangle. \tag{C.5}$$

Combining (C.4) and (C.5) shows that $U$ is unitary, i.e. for all $|g\rangle, |f\rangle \in \mathbb{H}$,

$$\langle g|U^\dagger U|f\rangle = \langle g|f\rangle. \tag{C.6}$$

**Theorem:** (Marshall Stone, 1932) Let $\{U(t): t \in \mathbb{R}\}$ be a set of unitary operators on a Hilbert space, $\mathbb{H}$, $U(t): \mathbb{H} \to \mathbb{H}$, such that $U(t+s) = U(t)U(s)$ and

$$\forall t_0 \in \mathbb{R}, |f\rangle \in \mathbb{H}, \lim_{t \to t_0} U_t|f\rangle = U_{t_0}|f\rangle$$

(i.e. $U$ is strongly continuous) then there exists a unique self-adjoint operator $H$ such that $U(t) = e^{-iHt}$.

*Proof:* The derivative of $U$ is

$$\dot{U} = \lim_{dt \to 0} \frac{U(t+dt) - U(t)}{dt} = \lim_{dt \to 0} \frac{U(dt)U(t) - U(t)}{dt}$$

$$= \left(\lim_{dt \to 0} \frac{U(dt) - 1}{dt}\right) U(t) = U(t) \left(\lim_{dt \to 0} \frac{U(dt) - 1}{dt}\right). \tag{C.7}$$

This prompts the definition of the Hamiltonian operator:

**Definition:** The **Hamiltonian operator** $H: \mathbb{H} \to \mathbb{H}$ is given by

$$H = i\left(\lim_{dt \to 0} \frac{U(dt) - 1}{dt}\right). \tag{C.8}$$

The Hamiltonian has no dependency on $t$. We have

$$\dot{U}(t) = -iHU(t) = -iU(t)H. \tag{C.9}$$

So $-iH = U^\dagger \dot{U} = \dot{U}U^\dagger$. Since $U$ is unitary, for a small time $dt$,

$$1 = U^\dagger(t+dt)U(t+dt) \approx [U^\dagger(t) + \dot{U}^\dagger(t)dt][U(t) + \dot{U}(t)dt]. \tag{C.10}$$

Ignoring terms in squares of $dt$, and using $-iH = U^\dagger \dot{U}$, $iH = \dot{U}^\dagger U$,

$$U^\dagger(t)U(t) - iH^\dagger dt + iH dt \approx 1. \tag{C.11}$$

Using unitarity of $U$, we find that $H$ is Hermitian, $H = H^\dagger$. (C.9) has solution,

$$U(t) = e^{-iHt}. \tag{C.12}$$

**Corollary:** The wave function satisfies the Schrödinger equation

$$\partial_0 f(t, x) = -iHf(t, x) \tag{C.13}$$

*Proof:* Differentiate the wave function using (C.9),

$$\partial_0 f(t, x) = \langle x|\dot{U}|f(0)\rangle = \langle x|-iHU(t)|f(0)\rangle = \langle x|-iH|f(t)\rangle = -iHf(t, x) \quad (C.14)$$

The Dirac equation is regarded as a relativistic Schrödinger equation. The argument must be adapted for photons, because the photon wave function refers to measurement of the position of the annihilation of a photon, not to a position eigenstate. For photons, the Schrödinger equation is replaced with the Gupta-Bleuler gauge condition and the massless Klein-Gordon equation, for which the solution is again a wave in a flat configuration space.

**Corollary:** Newton's first law.

*Proof:* For a non-interacting particle, $H = E$ = const where $E^2 = (p^0)^2 = m^2 + \mathbf{p}^2$ for some constant $m$. For an initial state $|f\rangle$ with momentum space wave function $\langle p|f\rangle$, the general solution is

$$f(x) = \left(\frac{1}{2\pi}\right)^{3/2} \int d^3p \langle p|f\rangle \, e^{-ix\cdot p} \quad (C.15)$$

Thus the momentum space wave function $\langle p|f\rangle$ does not change in time. It should be noted that the dot product, $x \cdot p$, in (C.15) uses Minkowski metric. It arises from the solution of a differential equation, *not* the physical metric of spacetime.

## Appendix D    Geodesic motion

A plane wave is a solution of the Schrödinger equation with $H = E$ = const. Thus momentum, $p$, does not change in time for a non-interacting particle. This is, in effect, Newton's first law. It applies in Minkowski tangent space, not in curved space-time. Quantum states are defined for particles in the neighbourhood of an observer on a synchronous slice at time $t$. In the classical correspondence there is sufficient information that states are effectively measured (effective measurement is defined in Francis 2013 & 2015 to mean that information is available in principle such that the result of a measurement is deterministic, whether or not a measurement is actually performed). To describe a continuous classical motion, we may take a collection of synchronous slices, and treat each slice as the final state of one quantum step and as the initial state of the next quantum step. We then allow the step size to go to zero, to obtain a foliation. At each stage of the motion, quantum theory is formulated using the non-physical Minkowski metric of tangent space.

Classical motion is deterministic and may be described as an ordered sequence, $|f_i\rangle = |f(t_i)\rangle$, of effectively measured states at instances $t_i$ such that $0 < t_{i+1} - t_i < \delta$ where $\delta$ is sufficiently small that there is negligible alteration in predictions in the limit $\delta \to 0$. Each state, $|f_i\rangle$, is a multiparticle state in Fock space, labelled according to mean properties determined by an effective classical measurement. For the motion between times $t_i$ and $t_{i+1}$, $|f_i\rangle$ may be regarded as the initial state and $|f_{i+1}\rangle$ may be regarded as the final state. Since the time evolution of mean properties is deterministic, there is no collapse and $|f_{i+1}\rangle$ is the initial state for the evolution to $t_{i+2}$.

Classical formulae are recovered by considering a sequence of initial and final states in the limit as the maximum time interval between them goes to zero. In each stage of evolution, momentum is parallel displaced. This gives the same result as parallel displacement in Minkowski tangent space. The cumulative effect of such infinitesimal parallel displacements is parallel transport in the direction of momentum. The same argument holds for a classical beam of light, in which each photon wave function is localised within the beam at any time, and for a classical field which has a measurable value at each point where it is defined (up to the possible addition of an arbitrary gauge function).